\documentclass[aps,pra,twocolumn,superscriptaddress,floatfix]{revtex4-2}   	

\usepackage{graphicx}
\usepackage{amssymb}
\usepackage{amsmath}
\usepackage{multirow}
\usepackage{braket}

\usepackage{array}
\newcolumntype{P}[1]{>{\centering\arraybackslash}p{#1}}

\begin{document}

\title{Elucidating the roles of collision energy and photon momentum transfer in the formation of ultralong-range Rydberg molecules}

\author{C. Wang}
\affiliation{Department of Physics and Astronomy, Rice University, Houston, TX  77005-1892, USA}
\affiliation{Smalley-Curl Institute, Rice University, Houston, TX  77005-1892, USA}
\author{Y. Lu}
\author{S. K. Kanungo}
\affiliation{Department of Physics and Astronomy, Rice University, Houston, TX  77005-1892, USA}
\author{F. B. Dunning}
\author{T.~C.~Killian}
\affiliation{Department of Physics and Astronomy, Rice University, Houston, TX  77005-1892, USA}
\affiliation{Smalley-Curl Institute, Rice University, Houston, TX  77005-1892, USA}
\author{S. Yoshida}
\affiliation{Institute for Theoretical Physics, Vienna University of Technology, Vienna, Austria, EU}

\begin{abstract}

Spectroscopic measurements of the rotational distribution of $^{84}$Sr and $^{86}$Sr $ ~5sns~^1S_0$ ultralong-range Rydberg molecular dimers created via photoassociation in a cold gas are reported.  The dimers are produced by two-photon excitation  via the 5s5p~$^1P_1$ intermediate state.  The use of singlet states permits detailed study of the roles that the initial atom-atom interaction, photon momentum transfer during Rydberg excitation, and sample temperature play in determining the spectral lineshape and final dimer rotational distribution.  The results are in good agreement with the predictions of a model that includes these effects.  The present work further highlights the sensitivity of ultralong-range Rydberg molecule formation to the state of the initial cold gas.
\end{abstract}

\maketitle 

\section{Introduction}

Interest in ultralong-range Rydberg molecules (ULRRMs), which comprise a Rydberg atom in whose electron cloud are embedded one (or more) weakly-bound ground-state atoms, has increased steadily over the years~\cite{srs18,fhs20,sdm16,let19,eil19}.  Such molecules, which represent a new molecular class, have provided a valuable microscale laboratory in which to study low-energy electron-atom scattering at energies not readily accessible using alternate techniques~\cite{sln16,edh19}, have furnished a powerful probe of non-local spatial correlations in cold quantum gases~\cite{wkd19,kld23}, illuminating the important role played by quantum statistics, and have, through  measurements of ULRRM formation in dense Bose-Einstein condensates, provided the opportunity to study many-body phenomena, such as the creation of Rydberg polarons~\cite{csw18,swd18,tco09,spe23}.

ULRRMs are bound by scattering of the Rydberg electron from the ground state atom which, as demonstrated in earlier studies, can frequently be described using a zero-range Fermi pseudo-potential~\cite{tle18,lwk22}.  The resulting molecular Born-Oppenheimer potential can support multiple vibrational states.  In the case of strontium $ns$ Rydberg states, the ground $v=0$ vibrational state is of particular interest because it is strongly localized near the outer classical turning point at an internuclear separation, $R_n \sim 1.8(n-\delta)^2 a_0$, where $a_0$ is the Bohr radius, and $\delta$ is the quantum defect. 

The vibrational structure of ULRRMs has been examined in a number of earlier studies.  Recent work, however, extended measurements to include the study of their rotational structure~\cite{lwk22}.  These studies employed $^{86}$Sr 5sns~$^3S_1$ dimers because the scattering length for $^{86}$Sr-$^{86}$Sr collisions is unusually large, $a_s = 811 a_0$, and is comparable in size to the internuclear separation for $v=0$ strontium ULRRM dimers with values of $n\sim 25$.  Since the value of $a_s$ defines the position of a node in the atom-atom scattering wavefunction, the probability of finding an atom pair with initial separations close to this value is reduced.  Thus, for values of $n$ near 25, the Franck-Condon overlap between the initial scattering state and the final ($v$=0) molecular state is reduced, thereby suppressing dimer formation via the $s$-wave channel. This, in turn, allows the effects of higher-partial-wave scattering, which can lead to the creation of rotationally-excited dimers, to be more visible. Furthermore, in the formation of ULRRM dimers only one atom in the initial ground-state pair acquires photon momentum through its excitation which, given the large size of the product dimer, can lead to significant angular momentum transfer to the atom pair. The earlier experimental results for $^3S_1$ dimers, however, were not in good agreement with the predictions of a theoretical model which incorporated the effects of both the scattering length and the recoil photon momentum.  The interaction of the electron spins with the molecular rotational motion was estimated to be small and was neglected. In this paper photoassociation spectra for Rydberg dimers created using singlet $5sns$~$^1S_0$ Rydberg states are analyzed to examine if the discrepancies observed in the previous study are peculiar to strontium triplet Rydberg states.

The $^{86}$Sr 5sns $^3S_1~(m=+1)$ Rydberg dimers employed in the earlier experiments~\cite{lwk22} were created by two-photon excitation via the intermediate 5s5p $^3P_1$ state using lasers operating at 689 and 320~nm.  The role of photon momentum transfer could not be clearly identified because, even when the laser beams counter-propagate, significant momentum transfer occurs.  Here we remove this limitation by studying the formation of 5sns~$^1S_0$ dimers created by two-photon excitation via the intermediate 5s5p~$^1P_1$ state, which requires lasers operating at 461 and 413 nm.  Thus, if the lasers counter-propagate, the net photon momentum transfer is very small, thereby providing a benchmark against which to identify the effects of photon momentum transfer.  To examine the effects of photon momentum transfer, co-propagating laser beams are employed.  

The effects of the initial atom-atom interaction are explored through comparative measurements using $^{84}$Sr for which the $s$-wave scattering length, $a_s = 123$$a_0$, is much less than that for $^{86}$Sr.  

The experimental measurements are compared with the results of model calculations that include the effects of both the initial atom-atom interaction and of photon momentum transfer.  The theoretical predictions are in good agreement with experiment and highlight the important role that these factors, together with the initial cold gas temperature, play in determining the rotational distribution of the product ULRRMs.

\section{Experimental method}

The experimental methods are described in detail elsewhere~\cite{ssk14,dmm09,dad15}.  Briefly, strontium atoms are laser-cooled to a few microKelvin and loaded into a  ``pancake''-shaped optical dipole trap (ODT) formed from two crossed elliptical 1064 nm laser beams.  Evaporative cooling is used to further lower and control the atom temperature.  The Rydberg molecules  are created by two-photon excitation via the intermediate 5s5p$^1P_1$ state.  The necessary radiation at 413 and 461~nm is provided by frequency-doubled diode laser systems that are stabilized to high-finesse ultralow-expansion optical cavities.  To limit trap loss and heating from photon scattering, the 461~nm laser was blue-detuned by $\sim18$~GHz. The photoexcitation lasers are typically applied for $\sim20~\mu$s.  Whereas the laser linewidths are small, typically a few kHz, the effective overall linewidth, measured by observing the excitation of atomic Rydberg states using counter-propagating beams, is larger $\sim50$ kHz, and is limited by transform broadening (when using counter-propagating laser beams Doppler broadening is minimal.)  It is, however, a significant improvement in linewidth as compared to that in earlier measurements, $\sim 120-140$~kHz \cite{lwk22}, and is sufficiently narrow as to  allow detailed spectroscopic studies of molecular photoassociation. 

  The ODT was loaded with $\sim 5-10\times 10^5$ atoms, resulting in trap densities of up to $\sim 1\times 10^{13}$~cm$^{-3}$.  
The final atom number, and temperature, were determined through absorption imaging on the $5s^2~^1S_0 \rightarrow 5s5p~^1P_1$ transition following release of the atoms from the ODT and a fall time of $\sim20$~ms.

The number of Rydberg molecules produced is determined through ionization in a pulsed electric field~\cite{sdu83,gal94}.  The resulting electrons are directed to a dual-microchannel-plate for detection and counted.  Experimental limitations on the size of the ionizing field that could be generated in the experimental volume limited measurements to states with $n\geq29$.  Excitation rates were kept low, typically $\lesssim 0.3$ per experimental cycle, to minimize possible effects due to Rydberg-Rydberg interactions.  The need to maintain a constant sample temperature and density during a set of measurements limited the number of experimental cycles that could be undertaken using a single cold-atom sample to $\sim 200$.

\section{Theoretical analysis}
\label{theoretical_analysis}

The present results are interpreted using a theoretical model that has been described in detail previously~\cite{lwk22}, and which includes the effects of ground-state atom-atom interactions in the initial scattering channel, together with the recoil momentum associated with photon momentum transfer.

Since a Rydberg dimer is formed by a weak perturbation to isolated atoms, the two atoms in the molecule can be treated as composite, identical bosonic particles as they are in the initial scattering state.  Thus, the Hamiltonian of the interacting atoms is invariant with an exchange of both electron and ion coordinates simultaneously, and the wavefunctions are symmetric with respect to the overall exchange. However, for an initial scattering state in which both atoms are in the same electronic state (the $5s^2$ ground state), the wavefunction becomes symmetric with respect to only the ion coordinates. When a partial-wave representation $\vert k, N^\prime,M^\prime_N\rangle$ is adopted for the relative motion of the colliding pair, where $N^\prime$ is the rotational quantum number, $ M_N^\prime$ its projection 
on the quantization axis, and $\hbar$k is the relative momentum, only even partial waves are allowed for the initial scattering state. Thus, when the effects of recoil photon momentum are negligible, only even, $N =0, 2, 4,...$, rotational states of the Rydberg dimer can be created.  Since the total orbital angular momentum of the valence electrons in strontium vanishes, the total mechanical angular momentum of the
product dimer is completely characterized by the quantum numbers $N$ and $M_N$.  The transition amplitude is given by an inelastic form factor that may be written
\begin{eqnarray}
&&F_{v,N,M_{N}}(k,N^{\prime},M_N^{\prime})
\nonumber \\
&& = \frac{1}{\sqrt{2}} \langle v,N,M_N\vert [e^{-i(\vec{\kappa}/2)\cdot\vec{R}}
+(-1)^N e^{i(\vec{\kappa}/2 )
\cdot \vec{R}}]\vert k,N^{\prime},M^{\prime}_N \rangle
\nonumber \\
\end{eqnarray}
where $v$ denotes the final vibrational state and $\vec\kappa$ is the sum of the wave vectors associated with the laser fields. Here we assume no mixing between vibrational and rotational levels for the ground vibrational level since the energy difference between the vibrational levels $v=0$ and 1 is typically much larger than the rotational energies. This form factor is similar to that used in molecular spectroscopy except that it includes the effects of photon recoil momentum.  The form factor is evaluated using spherical representations of both the initial asymptotic scattering state of the atomic pair and the final state of the Rydberg molecule whence the transition amplitude can be approximated by
    \begin{equation}
    F_{v,N,M_N}(k,N^{\prime},M^{\prime}_N)\propto j_{\lambda}(\kappa R_n/2)\tilde{j}_{N^{\prime}}(kR_n),
    \label{eq:fc0}
    \end{equation}
where $\lambda\simeq\vert N-N^\prime\vert$ is the change in rotational quantum number and
    \begin{equation}
    \tilde{j}_{N^{\prime}}(kR)
    =\left\{
    \begin{array}{ll}
    \frac{\sin[k(R-a_s)]}{kR}&N^{\prime}=0,\\
    j_{N^{\prime}}(kR)&N^{\prime}> 0
    \end{array}\right.
    \label{eq:Bessel}
    \end{equation}
where $j_{N^{\prime}}(kR)$ is the spherical Bessel function.  This form factor
resembles the Franck-Condon (FC) overlap between the  initial scattering state and final molecular state.  The Franck-Condon principle, however, argues that electronically- and
vibrationally-excited states are formed by ``vertical'' transitions at a given internuclear separation, and that rotational angular momentum will be conserved during photoassociation.  The non-diagonal transitions with $N-N^\prime\neq 0$ seen here result from the interplay of photon momentum recoil and the large molecular size.  The angular momentum transferred to the dimer, $\sim\kappa R_n/2$, when using counter-propagating beams, $\sim0.04$~a.u., is very much less than when using co-propagating beams, $\sim0.9$~a.u..  

The final excitation strength, $f(\omega)$, is obtained by averaging over the Boltzmann distribution of the relative momenta for atoms in the trap and the effective laser linewidths,       
    \begin{eqnarray}
    f(\omega)
    &\propto& \sqrt{\frac{2}{\pi}} \left(\frac{\hbar^2}{\mu k_BT}\right)^{3/2} 
    \sum_{v,N,M_N} \int dk k^2 e^{-h^2k^2/(2\mu k_BT)}
    \nonumber \\
    &&
    \times \sum_{N^\prime,M^\prime_N} \vert F_{v,N,M_N}(k, N^{\prime}, M_N^{\prime})\vert^2 L_{v,N}(k,\omega),
    \label{eq:final strength}
    \end{eqnarray}
where $L_{v,N}$ is a normalized line-shape function which, for the present experimental data, can be well approximated by a distribution including the Doppler broadening induced by the center of mass momentum $\hbar \vec{k}_{\rm CM}$ of the dimer,
    \begin{eqnarray}
    && L_{v,N}(k,\omega)
    \nonumber \\
    && = \frac{1}{\pi} \left(\frac{\hbar^2}{2\pi M k_BT}\right)^{3/2} 
    \int d\vec{k}_{\rm CM} e^{-h^2k_{\rm CM}^2/(2 M k_B T)} 
    \nonumber \\
    && \times
    \frac{2\Gamma}{4(\omega - \hbar \vec{\kappa} \cdot \vec{k}_{\rm CM} / M + \hbar k^2/(2\mu)-E_{v,N}/\hbar)^2 + \Gamma^2}
    \label{eq:Lorentzian}
    \end{eqnarray}
where $M$ is the total mass of the dimer and $\mu=M/4$ the reduced mass, $\hbar^2 k^2/2\mu$ the initial relative kinetic energy of the collision pair, $\omega$ the two-photon detuning from atomic resonance for an atom at rest, and $\Gamma$ the full-width-at-half-maximum (FWHM) of the effective instrumental linewidth.  The peak position in Eq.~\ref{eq:Lorentzian} is determined by the initial relative kinetic energy of the collision pair, i.e., $E_{v,N}\simeq \hbar\omega + \hbar^2k^2/(2\mu)$ on resonance, and energy conservation during excitation.  Note that thermal averaging, i.e., the convolution of Lorentzians with the Boltzmann distributions contained in Eqs.~\ref{eq:final strength} and \ref{eq:Lorentzian} leads to an increased linewidth and asymmetric non-Lorentzian overall line profiles~\cite{jlt99} that can be calculated using the molecular binding energies $E_{v,N}$ and the Franck-Condon overlap evaluated from the molecular Hamiltonian. 

In order to extract rotational energies $E_{v=0,N}$ and the contribution from each rotational level to the photoassociation spectra, the measured spectra are fit using the expression
\begin{eqnarray}
    f_{\rm fit}(\omega)
    &=& \left(\frac{2\hbar^2}{\pi \mu k_BT}\right)^{3/2} 
    \sum_{N} C_N 
    \nonumber \\
    && \times \int dk \, k^2 
    \frac{\vert {\cal F}_N(k)\vert^2~\Gamma e^{-h^2k^2/(2\mu k_BT)}}{4(\omega + \hbar k^2/(2\mu)-E_{v=0,N}/\hbar)^2 + \Gamma^2}
    \nonumber \\
    \label{eq:fit}
\end{eqnarray}
and the measured sample temperatures $T$ while treating the effective linewidth $\Gamma$, $E_{v=0,N}$, and weights, $C_N$, of the contributions from each rotational level as adjustable parameters. The Franck-Condon overlap is approximated using
Eq.~\ref{eq:fc0} as
\begin{equation}
{\cal F}_N(k) = \sum_{N'}  j_{\lambda}(\kappa R_n/2)\tilde{j}_{N^{\prime}}(kR_n)
\end{equation}
with $\lambda = |N-N'| = 0$ for even $N$ and 1 for odd $N$. In fitting the data for counter-propagating beams, i.e. in the limit of small $\kappa$,
it is further assumed that $j_{\lambda}(\kappa R_n/2) \sim \delta_{\lambda,0} = \delta_{N,N'}$. Sizable uncertainties in the fitted values, however, can result if the different contributions are not well-resolved.  As will be demonstrated, the excellent quality of the fits obtained using the predicted lineshapes, i.e., Eq.~\ref{eq:fit}, provides strong support for the present model.

\section{Results and discussion}
\begin{figure}
    \centering
        \includegraphics[width=1.0\linewidth]{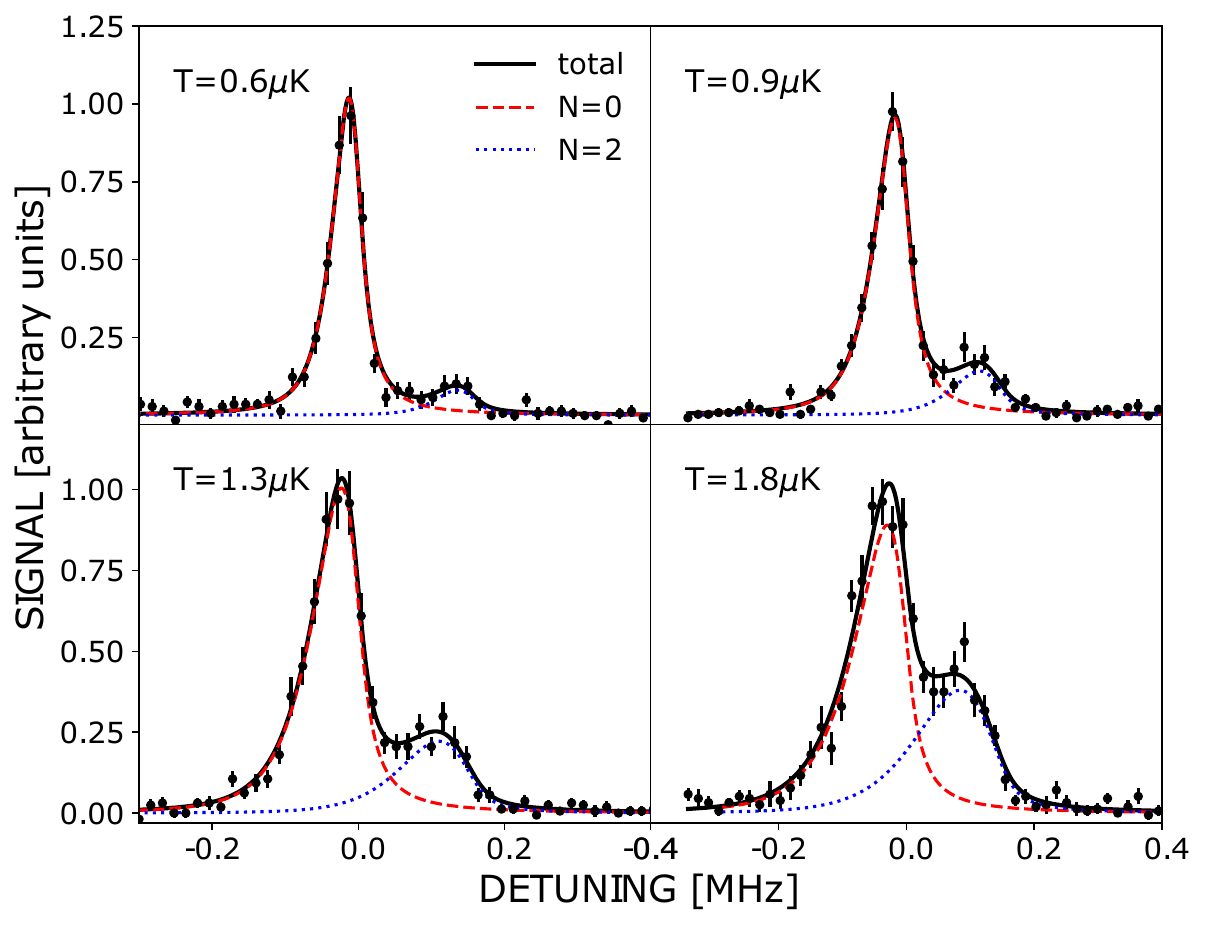}
        \label{fig:enter-label}
        \caption{\label{fig:spectra}Photoassociation spectra recorded when creating ground-state $v=0$  $^{86}$Sr $5s29s ^1$S$_0$ ULRRM dimers using counter-propagating laser beams and the sample temperatures indicated.  To emphasize the changes in spectral lineshape that occur, the results in each panel are normalized to the same peak height.  Fits obtained using the sum of contributions from just two states are included (see text).  Note that the horizontal axis in this, and later, figures is expressed in units of frequency, i.e., $\omega/2\pi$.
    }
\end{figure}

Figure~\ref{fig:spectra} shows photoassociation spectra recorded when creating 
 $^{86}$Sr $n=29$ $^1S_0$ dimers in the ground, $v=0$, vibrational state using counter-propagating laser beams and a number of representative sample temperatures.  To enable a more direct comparison of the spectral profiles, each data set is normalized to the same peak height.  (As the sample temperature is reduced, however, the number of atoms remaining in the trap is reduced, resulting in lower signal levels.)  As shown in Fig.~\ref{fig:spectra}, each measured spectrum can be well fit using Eq.~\ref{eq:fit}  and the sum of contributions from just two product states whose separations, $\Delta E$, are observed to be essentially independent of sample temperature. 
 
These observations point to the formation of just two
 \begin{table}
     \begin{tabular}{|c|c|c|c|c|c|}
     \hline
     Rydberg&levels&\multicolumn{4}{c|}{level spacings, kHz}\\
     \cline{3-6}
     state&&A&B&C&D\\
     \hline
     5s29s$^1S_0$&N=0 to N=2&166&177&167$\pm$12&173$\pm$12\\
     \hline
     5s30s$^1S_0$&N=0 to N=2&142&152&140$\pm$11&141$\pm$11\\
     \hline
     5s31s$^1S_0$&N=0 to N=2&122&132&127$\pm$10&123$\pm$10\\
      \hline
      5s32s$^1S_0$&N=0 to N=2&105&114&106$\pm$9&113$\pm$9\\
      \hline
      5s33s$^1S_0$&N=0 to N=2&92&100&94$\pm$8&92$\pm$8\\
      \hline
     5s29s$^1S_0$*&N=0 to N=1&55&59&54$\pm$7&\\
     \cline{2-6}
     &N=0 to N=2&166&177&179$\pm$25&\\
     \hline
     \end{tabular}
     \caption{\label{Ta:measured} Calculated and fitted level spacings for the transitions indicated.  The asterisk denotes measurements using co-propagating rather than counter-propagating, laser beams.  Column A shows rotational level spacings derived from the eigenenergies $E_{v=0,N}$ of the Rydberg dimer Hamiltonian. Column B lists the approximate values obtained when assuming the molecule behaves as a rigid rotor (Eq.~\ref{eq:levels}).  Columns C and D show level separations deduced from fits to measured $^{86}$Sr and $^{84}$Sr spectra, respectively, and represent the average of measurements taken at several different sample temperatures and trap operating conditions.
     }
 \end{table}
 different final molecular states.  The energy separations, $\Delta E$, extracted from fitting the measured data are shown in Table~\ref{Ta:measured}.  Also included in Table~\ref{Ta:measured} are the calculated values of $\Delta E$ derived from the eigenenergies of the Rydberg dimer Hamiltonian together with those predicted by assuming the molecule behaves as a rigid rotor, in which event the rotational levels can be approximated by
\begin{equation}
   \label{eq:levels}
   E_{v,N}
   \simeq E_{v,N=0} + \frac{\hbar^2 N(N+1)}{2\mu R_n^2}
   \end{equation}
where $R_n$ is the internuclear separation, $\sim 1.8(n-\delta)^2 a_0$, for the ground vibrational state. As seen in Table~\ref{Ta:measured}, the separations in the calculated eigenenergies between the $N=0$ and $N=2$ levels are very similar to those seen in the experimental measurements, pointing to the creation of $N=0$ and $N=2$ rotational states. We note that the separation between the peaks in the photoassociation spectra that correspond to the production of $N=0$ and 2 states is typically smaller by an amount comparable to the thermal energy, $k_BT \sim 20-40$~kHz, as compared to their actual energy separations (see Table~\ref{Ta:measured}) due to the thermal averaging.   

\begin{figure}
    \centering
    \includegraphics[width=0.9\linewidth]{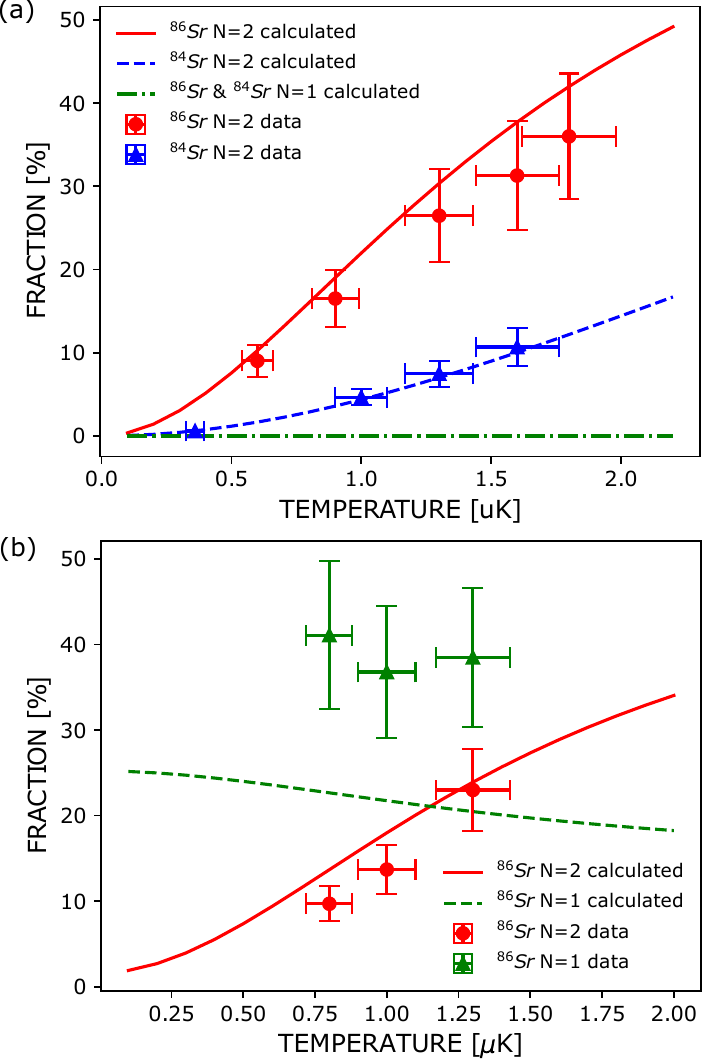}
\caption{\label{fig:temperature}
Model predictions of the temperature dependence of the fractional contributions from $N=1$ and 2 states to the total Sr $n=29 ^1S_0$ dimer signal for both (a) counter- and (b) co-propagating laser beams. (a) includes results for both $^{84}$Sr and $^{86}$Sr dimers, (b) results for only $^{86}$Sr dimers.  Values obtained from fits to the experimentally-measured spectra are also shown (symbols).
} 
\end{figure}

Figure~\ref{fig:temperature} shows, for 
$n=29$, the fractional contributions from $N=1$ and 2 states to the total dimer signal as a function of sample temperature when using co- and counter-propagating laser beams.  Values derived from fits to the measured spectra, obtained by integrating their separate contributions to the total signal, are shown by symbols.  The model predictions (lines) were calculated using the numerically-integrated form factor (Eq.1).  The calculated fractional contributions from $N=2$ states increase steadily with sample temperature reflecting the fact that the contributions from $N=2$ partial waves increase with increasing collision energy in this temperature range (Eq.~\ref{eq:fc0}). 

For counter-propagating beams (Fig.~\ref{fig:temperature}(a)) the model predictions are in good agreement with the measured values. The small discrepancies seen at high temperatures for $^{86}$Sr may be partly due to the fitting error as the overlap between the $N=0$ and 2 features becomes larger. No significant production of $N=1$ states is predicted, or seen,  consistent with the near-zero net photon momentum transfer.

\begin{figure}
     \centering
     \includegraphics[width=1\linewidth]{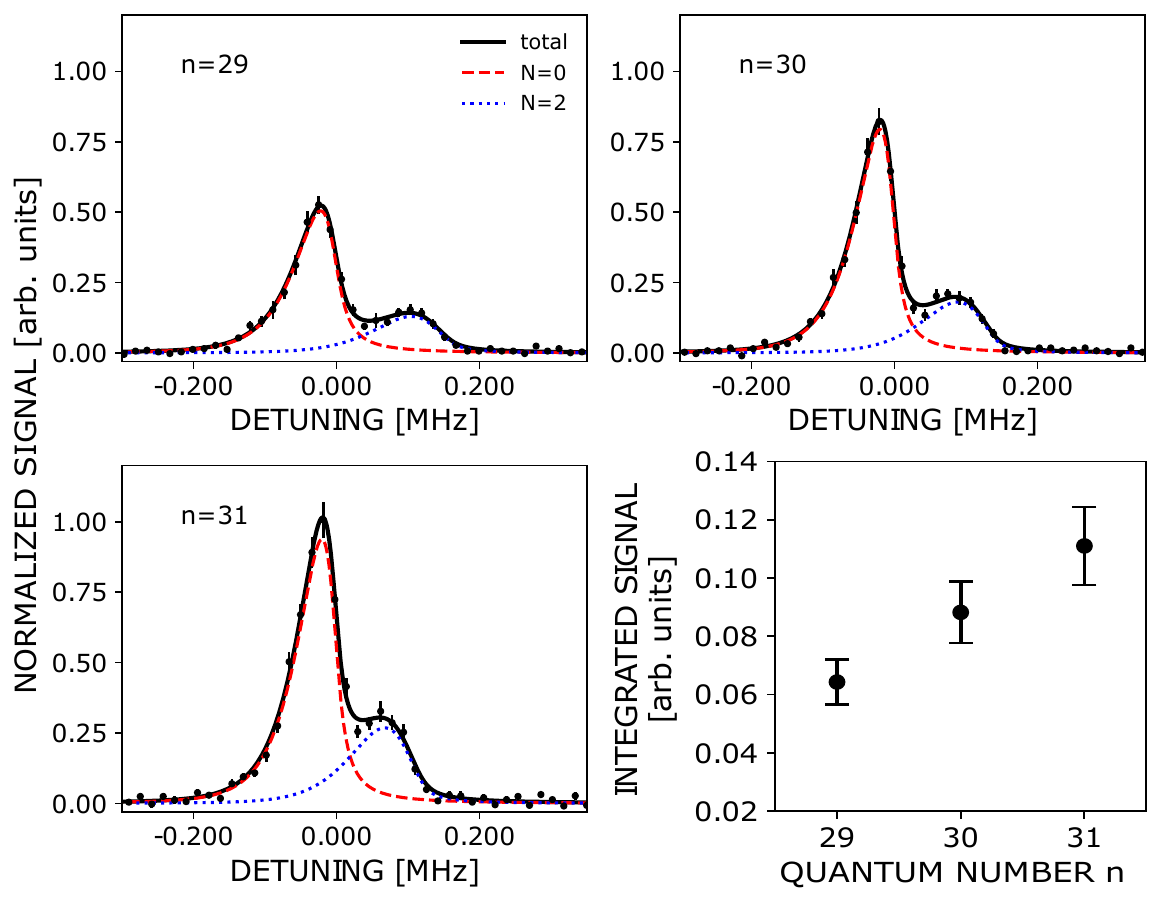}
     \caption{\label{fig:photoassociation}
     (a)-(c) Photoassociation spectra recorded when creating ground-state $v$=0 $^{86}$Sr $n=29$ to 31 $^1S_0$ dimers using counter-propagating laser beams and a sample temperature of $\sim 1.3\mu$K.  The results are normalized for small changes in trap density, laser powers, and oscillator strengths.  Fits obtained using the sum of contributions from just two states (see text), are also shown.  (d) $n$-dependence of the total dimer signal.
     }
\end{figure}

\begin{figure}

     \centering
     \includegraphics[width=0.9\linewidth]{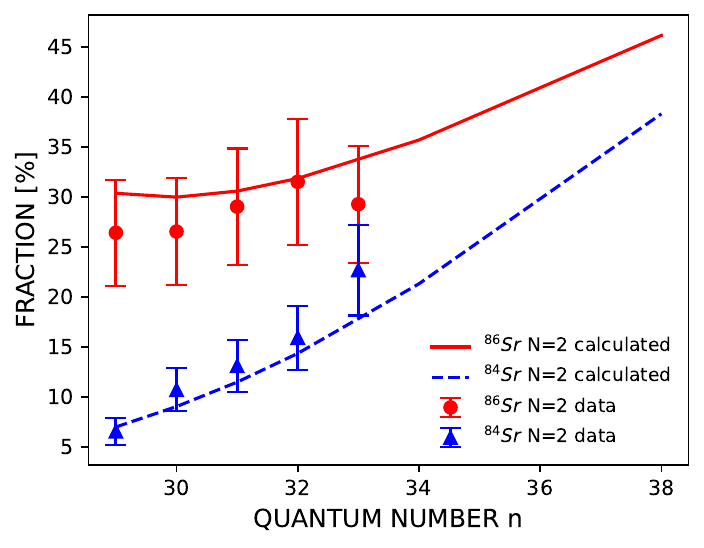}
\caption{\label{fig:fractional}
  Model predictions of the fractional contribution of $N=2$ states to the total dimer signal as a function of $n$ for both $^{84}$Sr and $^{86}$Sr together with the results of experimental measurements.  The results are for counter-propagating laser beams and a sample temperature of $1.3~\mu$K.
 }
 \end{figure}

Figure~\ref{fig:photoassociation} compares photoassociation spectra recorded when creating $n=29$ to 31 $^{86}$Sr 
dimers at a fixed sample temperature (1.3~$\mu$K).  These data were recorded on a single day to ensure similar trap conditions and are normalized for small variations in trap density, laser powers, and the $n$-dependence in the oscillator strengths for their creation~\cite{wkd19}.  As $n$ decreases towards $n=25$ (for which $R_n$ nearly matches $a_s$), the overlap between the atomic scattering wavefunction and the the ground-state molecular vibrational wavefunction decreases. Since for the present sample temperatures $s$-wave scattering typically dominates, this reduced overlap leads to a decrease in total dimer signal.  As seen in Fig.~\ref{fig:photoassociation}, the  (normalized) total dimer signal does decrease significantly as $n$ decreases from $n=31$ to $n=29$.  (Experiment shows that the dimer signal continues to grow as $n$ increases beyond 31.)  Figure~\ref{fig:photoassociation} also includes two-component fits to the data.  The resulting level separations are included in Table~\ref{Ta:measured} and are in very good agreement with the model predictions.  As $n$ increases, the separation between the rotational levels also decreases, ($\hbar^2 N(N+1)/(2 \mu R_n^2) \sim n^{-4}$), and for $n \geq 34$ their associated features can no longer be reliably resolved.  Note that, as the sample temperature rises, so too does the fractional contribution of $N=2$ states to the total dimer signal.  Figure~\ref{fig:fractional} 
shows the calculated fractional contribution from $N=2$ states as a function of $n$ for a sample temperature of $1.3~\mu$K together with the fitted values, which are again in reasonable agreement with model predictions. 

\begin{figure}
    \centering
    \includegraphics[width=1\linewidth]{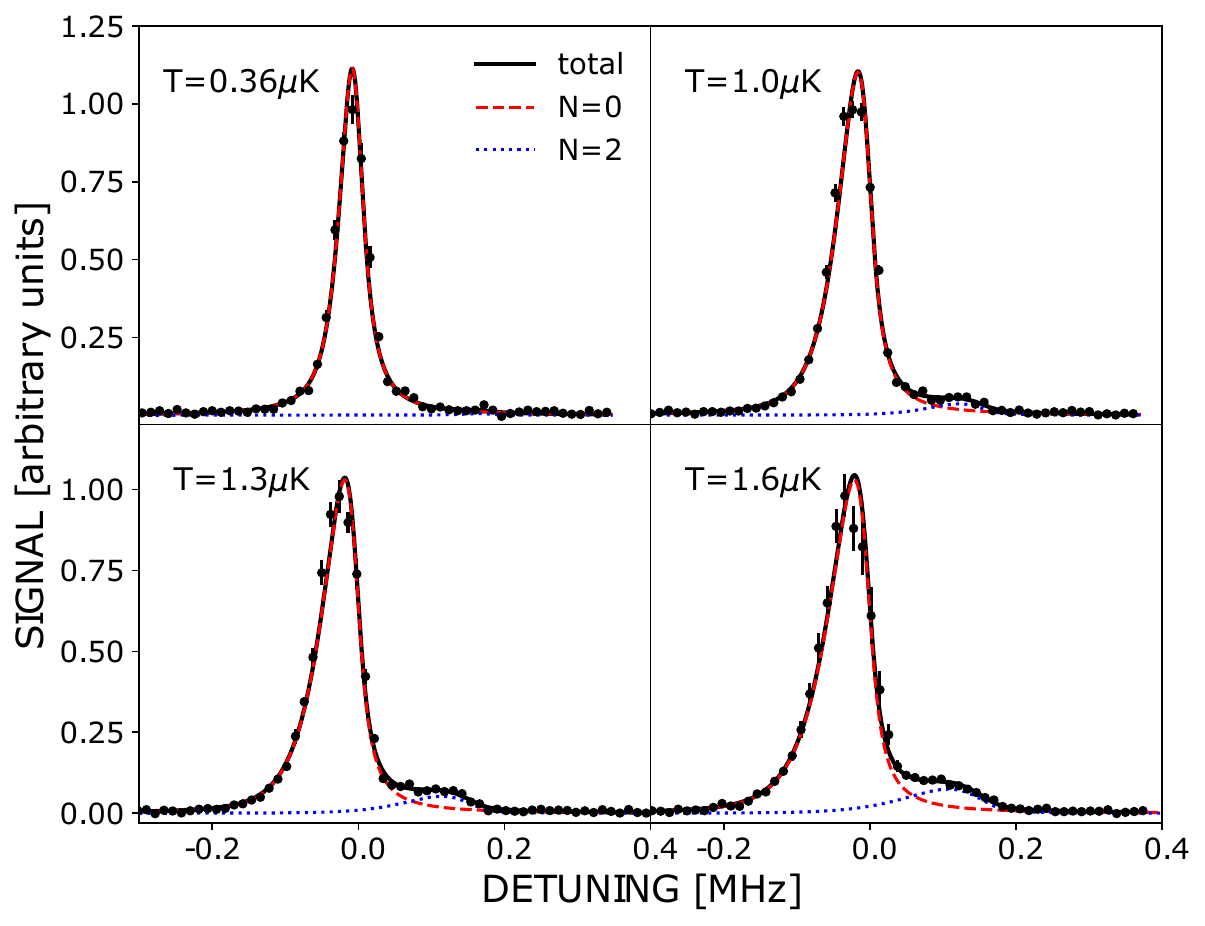}
    \caption{Photoassociation spectra recorded when creating $^{84}$Sr $^1S_0$ $v$=0 5s29s dimers using counter-propagating laser beams and the sample temperatures indicated.  Fits to the experimental data are also included.  To emphasize the changes in the spectral profiles that occur, the results are normalized to the same peak height.  
    }
    \label{fig:counter}
\end{figure}

To further illustrate the importance of $s$-wave suppression in the case of $^{86}$Sr dimers, Fig.~\ref{fig:counter} shows photoassociation 
spectra recorded when creating $^{84}$Sr ULRRM dimers.  For $^{84}$Sr, the $s$-wave atom-atom scattering length is small, $a_s\sim 123 a_0$, and there is no node in the scattering wavefunction at the internuclear separations explored in the present work.  In consequence, dimer production via the $s$-wave scattering channel is no longer suppressed.  However, little change is expected in the higher-partial-wave channels between $^{84}$Sr and $^{86}$Sr due to the presence of a long-range centrifugal barrier.  For $^{84}$Sr, therefore, $s$-wave scattering will be much stronger as compared to higher-partial-wave scattering.  The measurements do, however, reveal a small contribution to the total $^{84}$Sr dimer signal at a detuning that is consistent with that for creation of $N=2$ states and that increases with sample temperature. 
 
 As seen in Fig.~\ref{fig:fractional}, measurements show that the fractional contribution of $N=2$ states to the total dimer signal also increases as $n$ increases, behavior that is mirrored in the model predictions. 
 The agreement between theory and experiment is again very good, further validating the present model by demonstrating it produces reliable predictions even when considering an atom pair with very different $s$-wave scattering characteristics.  
 
 Given that the trap operating conditions for $^{84}$Sr are rather different to those for $^{86}$Sr, it is difficult to normalize data sets obtained for each species to one another to compare their absolute relative dimer production rates.  Nonetheless, attempts at such normalization suggest that, for $n=29$, the dimer production rate for $^{84}$Sr is some ten times larger than that for $^{86}$Sr.  As $n$ increases, however, this difference is seen to decrease due to the decreasing $s$-wave suppression for $^{86}$Sr.

 Consider now the case of co-propagating laser beams for which photon momentum transfer is maximized.
 \begin{figure}
     \centering
     \includegraphics[width=1\linewidth]{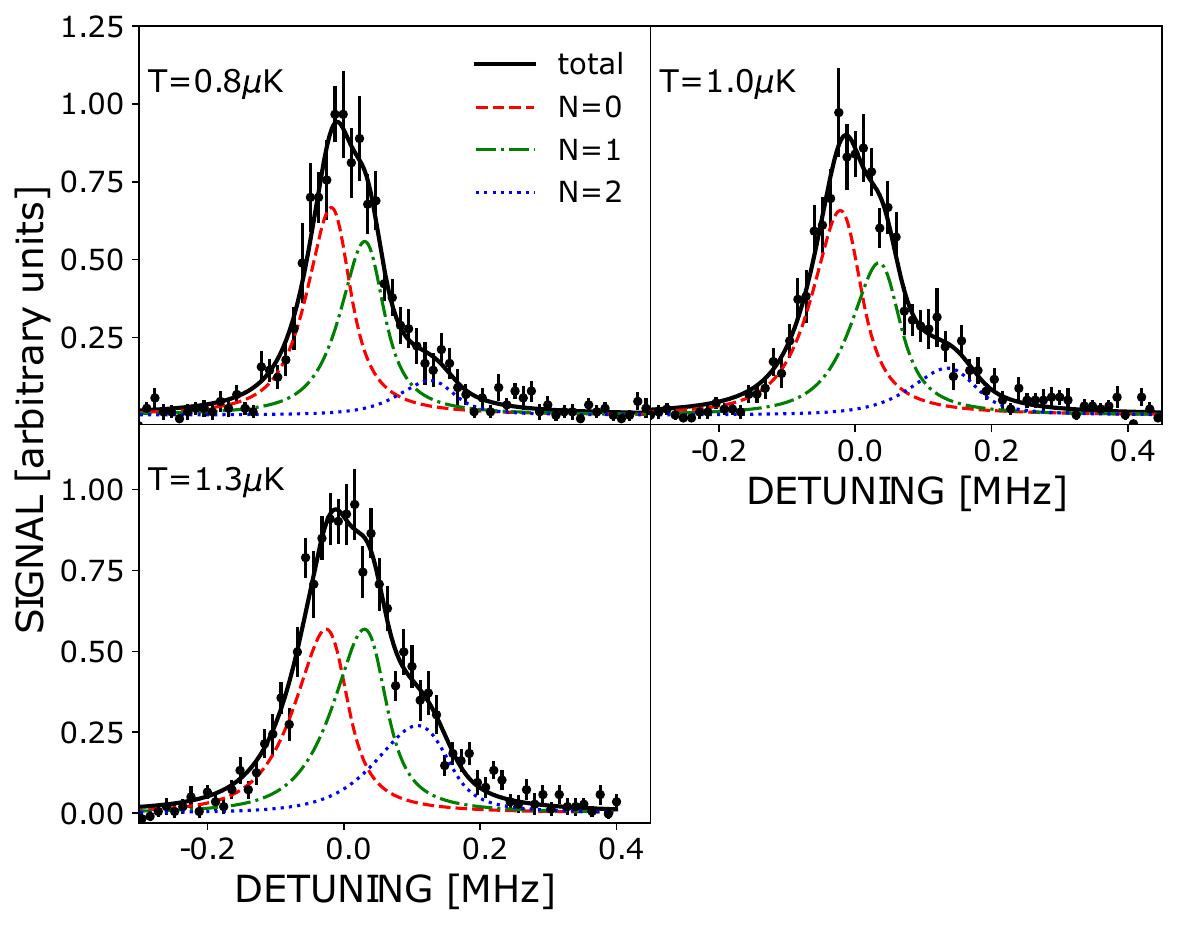}
     \caption{
     Photoassociation spectra recorded when creating $^{86}$Sr $n=29~^1S_0$ dimers using co-propagating laser beams and the sample temperatures indicated.  Also included are fits obtained using the sum of three states.
     }
     \label{fig:transfer}
 \end{figure}
Figure~\ref{fig:transfer} shows photoassociation spectra recorded using co-propagating laser beams when creating $n=29~^1S_0$ dimers.  Whereas for counter-propagating beams Doppler broadening should be very small, that for co-propagating beams will be significant, which is reflected in the increased widths of the spectral features.  
As a result the spectral features are less well resolved than for counter-propagating beams.  Attempts to obtain a good fit to the measured spectra assuming contributions from just two product states with separations corresponding to that expected between the  $N=0$ and $N=2$ levels proved unsuccessful.  However, as demonstrated in
Fig.~\ref{fig:transfer}, the spectra can be well fit by the sum of three separate contributions.  The separations between these three features are included in Table~\ref{Ta:measured} and are consistent with those predicted between the $N=0$ and $N=1$ and 2 features, pointing to the formation of a mixture of all three.   

Figure~\ref{fig:temperature} includes the results of model simulations of the fractional contributions from the $N=1$ and 2 rotational levels for co-propagating beams.  The predicted fractions of $N=2$ states, which are reasonably well resolved, are in good agreement with experimental observations.  The relative production  of $N=1$ states inferred from the fits to the data, however, is somewhat higher than suggested by the model (see Fig.~\ref{fig:temperature}), which might be attributed to uncertainties introduced by the considerable overlap of the $N=0$ and $N=1$ features. Nonetheless, even though these features are not well resolved, the quality of the fits is such that the $N=1$ contribution and position (see Table~\ref{Ta:measured}) are reasonably well determined.  Clearly, $N=1$ states provide a sizable contribution to the dimer signal demonstrating that photon momentum transfer is very important in the creation of rotationally-excited dimers.  

Earlier studies using $n=29~^3S_1$ dimers and co-propagating laser beams (which results in the same total photon momentum transfer as for $^1S_0$ dimers) suggested the additional production of $N=3$ dimers~\cite{lwk22}.  However, no evidence of their creation (at an expected detuning of $\sim350$~kHz) was seen in the present study of $^1S_0$ dimers.  Furthermore, the present model (see Fig.~\ref{fig:temperature}), does not predict the formation of significant numbers of such states. This suggests that the electron spin angular momentum might play a significant role in determining the final rotational distribution. However, direct coupling between the electron spin $S$ and the rotational motion $N$ is estimated be negligible. Further work, therefore, will be required to examine any other mechanisms by which such spin to rotational angular momentum transfer might occur.


\section{Conclusions}
The present measurements demonstrate the important roles played by atom-atom interactions, photon momentum transfer, and sample temperature in the production of rotationally-excited ULRRMs and validate the model described in Section~\ref{theoretical_analysis} which includes their effects.  In future studies with $^{86}$Sr it will be interesting, for example, to see if similar rotational structure can be resolved when creating the less-well-localized vibrationally-excited states, or even ground-state trimers, as well as exploring effects associated with reduced system mass by creating heteronuclear dimers in a two-component cold gas containing $^{86}$Sr and another alkali or alkaline earth element.  In addition, direct comparisons to detailed measurements using $^{86}$Sr $n^3S_1$ dimers will help unravel the role played by spin angular momentum in determining the final distribution of rotational states.

\begin{acknowledgments}
    Research supported by the National Science foundation under Grants No. PHY-1904294 and PHY-2110596, and the FWF(Austria) under Grant No. FWF-P35539-N, and Doctoral College FWF W 1243(Solids4fun).
\end{acknowledgments}
\bibliography{bibliography}
\end{document}